\documentclass[aps,pra,floatfix,notitlepage,showpacs]{revtex4-1}

\usepackage{graphicx}
\usepackage{amsfonts}
\usepackage{amsmath}
\usepackage{amssymb}
\usepackage{dsfont}
\usepackage{indentfirst}
\usepackage{dcolumn}
\usepackage{color}
\usepackage{bm}

\newcolumntype{d}[1]{D{.}{.}{#1}}

\renewcommand{\Re}[1]{\hbox{\rm Re}\left(#1\right)}         
\newcommand{\e}{\,\hbox{\rm e}}                                      

\begin{document}

\title{Slow light with three-level atoms in metamaterial waveguides}

\author{Benjamin R. Lavoie}
\email{brlavoie@ucalgary.ca}
\author{Patrick M. Leung}
\author{Barry C. Sanders}
\email{sandersb@ucalgary.ca}
\affiliation{Institute for Quantum Science and Technology, University of Calgary, Alberta, T2N 1N4, Canada}
\begin{abstract}

Metamaterial is promising for enhancing the capability of plasmonic devices. We consider a cylindrical waveguide with three-level $\Lambda$ atoms embedded in the dielectric core. By comparing metal cladding vs metamaterial cladding of a waveguide with $\Lambda$ atoms in the core, we show that, for a fixed amount of slowing of light due to electromagnetically induced transparency, the metamaterial cladding outperforms in terms of the inherent loss.


\end{abstract}


\maketitle


\section{Introduction}

Plasmonic devices show promise in increasing the speed of electronic devices and networks while still meeting the size requirements of modern electronic devices due to their ability to confine light to subwavelength scales~\cite{Brongersma:2010}. Combining such devices with optical phenomena, such as slow light, could yield an array of new optical devices to augment or replace existing technology. The feasibility of using nonlinear metamaterials for slow light has been considered~\cite{D'Aguanno:2008,Tassin:2009}, and a scheme for photon-echo quantum memory for surface plasmon--polaritons on a metamaterial interface has been proposed~\cite{Moiseev:2010book}. Controllable slow light is useful for optical delay lines, optical buffers~\cite{Boyd:2006}, and enhanced nonlinear interactions~\cite{ZBWang:2006}; however, achieving highly confined slow light with weak fields is impractical due to the high losses suffered as a result of the plasmonic confinement of light~\cite{Dionne:2006}. There exists a scheme~\cite{Kamli:2008,Moiseev:2010} using a flat interface metamaterial-dielectric waveguide for slowing highly confined light through electromagnetically induced transparency (EIT)~\cite{Lambropoulos:2007}, while using metamaterial to minimize losses. EIT with weak fields is interesting in its own right for fundamental reasons, to distinguish it from Autler-Townes splitting, and also for applications to weak-field sensing~\cite{Anisimov:2011}. However, the scheme has the limitation that field confinement in the waveguide is only in one transverse spatial dimension, which allows propagating fields to diverge. 

\begin{figure}[t,b] 
      \centering
	\includegraphics[width=.6\textwidth]{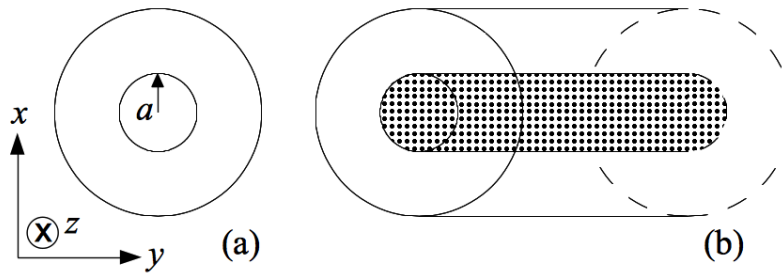}
	\caption[The cylindrical waveguide.]{(a) The cylindrical waveguide cross section. (b) A schematic of the metamaterial-dielectric waveguide embedded with three-level $\Lambda$ atoms (black dots) in the core. The outer layer (cladding) is a metamaterial whereas the core is a dielectric.}
     \label{fig:guidediagram}   
\end{figure}

\begin{figure}[t,b] 
       \centering
	\includegraphics[width=.4\textwidth]{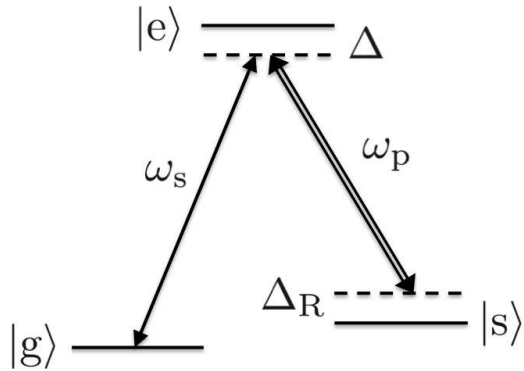}
	\caption{A schematic of the three-level $\Lambda$ atom. The ground state is $\left|\text g\right>$, the excited state is $\left|\text e\right>$, and $\left|\text s\right>$ is a metastable state. The frequency of the pump field is $\omega_{\text p}$ and of the signal field is $\omega_{\text s}$; and $\Delta$ and $\Delta_{\text R}$ are detunings.}
     \label{fig:EITsys}   
\end{figure}

We propose using a cylindrical waveguide structure as shown in Fig.~\ref{fig:guidediagram}, rather than a flat interface, to inherently provide confinement in both transverse directions and prevent field divergence. The waveguide is composed of a dielectric core and a metamaterial cladding that surrounds the core. Three-level $\Lambda$ atoms (Fig.~\ref{fig:EITsys}) are homogeneously embedded throughout the core of the waveguide and are driven by a pump field in a low-loss surface mode~\cite{Moiseev:2010}, thereby enabling EIT for slowing the signal field in another low-loss surface mode. To prepare EIT the atoms are initially strongly pumped to depopulate the $\left|\text s\right>$ level and relax to the $\left|\text g\right>$ state. This process creates Fano interference in the $\left|\text{g}\right>\leftrightarrow\left|\text{e}\right>$ transition, resulting in a transparency window at the $\left|\text{g}\right>\leftrightarrow\left|\text{e}\right>$ transition frequency. As a result of the refractive index change, the group velocity of the signal pulse is reduced when traveling through the transparent medium. After preparing for the EIT, the intensity of the pump field is lowered because the weaker the pump field, the greater the reduction of group velocity of the signal field. Hence, coherently controlled slow light is achievable by tuning the intensity of the pump field. Furthermore, by comparing the metamaterial-clad waveguide to a metal-clad one with the same permittivity, we show that a low-loss surface mode~\cite{Lavoie:2012} of the metamaterial-clad guide enables the same degree of slowing of light but with reduced losses. Our analysis is important to fabricating coherently controlled low-loss slow light devices with current metamaterial technology, and we provide conditions necessary for the existence of the low-loss surface mode for slowing light in the waveguide. 



In Sec.~\ref{theory}, we develop the theory for field propagation in the cylindrical metamaterial-clad waveguides without the doping of the three-level atoms. Then we extend the theory to allow for doping the atoms in the core of the waveguide. In Sec.~\ref{results}, we describe the numerical method we use to solve Maxwell's equations in the doped waveguide and present results obtained from the solutions. A discussion of our scheme, along with some considerations for implementation and possible applications, is given in Sec.~\ref{discussion}, and a summary of the work is given in Sec.~\ref{summary}.

\section{Theory}\label{theory}

Metamaterials are designed with artificial structure, giving them electromagnetic responses different to those of their constituent materials. We consider a fishnet-type metamaterial as they have been well studied, both experimentally and theoretically~\cite{Ramakrishna:2005,Boardman:2005,Boltasseva:2008,Xiao:2009}. Metamaterials of this design are inherently lossy, so the permittivity and permeability expressions must necessarily include dissipation terms. The permittivity of the fishnet metamaterial is described by the Drude model~\cite{Balmain:2005},
\begin{equation}
\frac{\epsilon_{\text{MM}}(\omega)}{\epsilon_{0}}=\epsilon_{\text{b}}-\frac{\omega_{\text{e}}^2}{\omega\left(\omega+i\Gamma_{\text{e}}\right)},\label{epsilonMM}
\end{equation}
with $\epsilon_{0}$ the permittivity of free space, $\epsilon_{\text{b}}$ the background permittivity, $\omega_{\text{e}}$ the plasma frequency of the metal, and $\Gamma_{\text{e}}$ the electric damping term. The permeability, which is in practice achieved through the structure of the metamaterial, is described by the relation~\cite{Penciu:2010}
\begin{equation}
\frac{\mu_{\text{MM}}(\omega)}{\mu_{0}}=\mu_{\text{b}}+\frac{F \omega^2}{\omega_0^2-\omega\left(\omega+i\Gamma_{\text{m}}\right)},\label{muMM}
\end{equation}
with $\mu_{0}$ the permeability of free space, $\mu_{\text{b}}$ the background permeability, $F$ a geometric parameter, $\omega_0$ the resonance frequency, and $\Gamma_{\text{m}}$ the magnetic damping term. Equations (\ref{epsilonMM}) and (\ref{muMM}) describe the electromagnetic properties of the metamaterial cladding and are used, along with a constant permittivity for the core, to determine the modes of the undoped metamaterial-dielectric waveguide.

After preparation for EIT, the $\left|\text s\right>$ state is not populated and the atoms are no longer excited by the pump field, which means the pump field effectively travels in the modes of the undoped waveguide. On the other hand, the signal field can excite the atoms, so the signal field travels in the modes of the doped waveguide. In the following, we develop the theory for finding the modes in the undoped waveguide before finding the modes of the doped waveguide.


The guided modes in a cylindrical waveguide are transverse electric (TE), transverse magnetic (TM), or a combination of the two (denoted HE or EH)~\cite{Yeh:2008}. The TM and TE modes are those with field components $H_{z}=0$ and $E_{z}=0$, respectively, with $z$ the propagation direction. We restrict our analysis to TM modes, as they are invariant in azimuthal angle $\phi$, which allows us to simplify the calculations. In general, the analysis can be done with any choice of modes.

For the TM modes of the metamaterial-dielectric waveguide, the electric fields have the form
\begin{equation}
\tilde{\bm E}(r,\phi,z,t)={\bm E}(r,\phi)\e^{i(\tilde{\beta} z-\omega t)}+\text{c.c.},
\end{equation}
with c.c.\ indicating the complex conjugate, and $\tilde{\beta}=\beta+i\alpha$ the complex propagation constants for the allowed modes of the waveguide, which are obtained by solving the dispersion relation. Using the definitions
\begin{equation}
J'_m\left(a\kappa\right):=\left.\frac{\partial J_m\left(r \kappa\right)}{\partial r}\right|_{r=a}
\end{equation}
and
\begin{equation}
K'_m\left(a\gamma\right):=\left.\frac{\partial K_m\left(r \gamma\right)}{\partial r}\right|_{r=a}{\textstyle ,}
\end{equation}
with $ J_m$ and $K_{m}$ the Bessel function and modified Bessel function, respectively, the dispersion relation for TM modes in an undoped cylindrical metamaterial-dielectric waveguide is~\cite{Yeh:2008}
\begin{equation}
\frac{\epsilon_{\text{d}}}{\kappa^{2}}\frac{J'_m\left(a \kappa\right)}{J_{m}\left(a\kappa\right)}+\frac{\epsilon_{\text{MM}}}{\gamma^{2}}\frac{K'_m\left(a\gamma\right)}{K_m\left(a\gamma\right)}=0\label{cyldisprel},
\end{equation}
with
\begin{equation}
\gamma:=\sqrt{\tilde{\beta}^{2}-\omega^{2}\epsilon_{\text{MM}}\mu_{\text{MM}}},\quad\kappa:=\sqrt{\omega^{2}\epsilon_{\text d}\mu_{\text{d}}-\tilde{\beta}^{2}},
\end{equation}
$\epsilon_{\text d}$ the permittivity of the dielectric core, and $a$ the core radius. With this dispersion relation, we can find the low-loss TM surface mode for the pump field as detailed in~\cite{Lavoie:2012}.

The susceptibility of the undoped waveguide core is simply the dielectric susceptibility,
\begin{equation}
\chi_{\text{d}} =\epsilon_{\text{d}}/\epsilon_{0}-1.
\end{equation}
When the core is doped with the three-level atoms, the susceptibility of the core becomes the sum $\chi_{\text{d}}+\chi(\omega,r)$, which alters the guided modes near the EIT resonance frequency from those of the undoped guide. To find the low loss TM surface mode of the signal field, we must solve the wave equation with the susceptibility of the doped waveguide core.


The susceptibility of the pumped three-level $\Lambda$ atoms under EIT is~\cite{Lambropoulos:2007}
\begin{equation}
\chi(\omega, r)=\frac{2a_{0}c}{n_{\text d}\omega}\frac{i\gamma_{\text{eg}}}{\gamma_{\text{eg}}-i\Delta
+\left|\Omega_{\text p}(r)\right|^{2}\left(\gamma_{\text{sg}}-i\Delta_{\text R}\right)^{-1}},\label{EITchi}
\end{equation}
with $c$ the speed of light in vacuum, $n_{\text d}=c\sqrt{\epsilon_{\text d}\mu_{\text d}}$ the refractive index of the dielectric core with $\mu_{\text d}$ the permeability of the dielectric core, $\gamma_{ij}$ the decay rate from level $i$ to level $j$ and
\begin{equation}
a_{0}:=\frac{3\pi c^{2}}{n_{\text d}^{2}\omega^{2}}\rho_{\text a}\approx\frac{3\pi c^{2}}{n_{\text d}^{2}\omega_{\text{eg}}^{2}}\rho_{\text a}
\label{eq:ndensity}
\end{equation}
with $\rho_{\text a}$ the number density of three-level atoms. The approximation in Eq.~(\ref{eq:ndensity}) is valid near the EIT resonance frequency. Defining $\omega_{ij}$ as the transition frequency for the $\left|i\right>\leftrightarrow\left|j\right>$ transition, the detunings are $\Delta=\omega_{\text{eg}}-\omega_{\text s}$, $\Delta_{\text R}:=\Delta-\omega_{\text{es}}+\omega_{\text p}$. The pump is set on resonance with the $\left|\text{s}\right>\leftrightarrow\left|\text{e}\right>$ transition such that $\Delta_{\text R}=\Delta$. The Rabi frequency of the pump is
\begin{equation}
\Omega_{\text p}(r)=\frac{1}{\hbar}{\bm d}\cdot {\bm E}_{\text{p}}(r),\label{rabifreq}
\end{equation} 
with $\bm d$ the dipole moment of the $\left|\text{s}\right>\leftrightarrow\left|\text{e}\right>$ transition.

Equations~(\ref{EITchi}) and (\ref{rabifreq}) have $r$ dependence but no $\phi$ or $z$ dependence. This is because the pump is in a TM mode, which has a radial dependence but no azimuthal dependence, implying $\phi$ invariance. Furthermore, the pump is approximated as nondepleting, which implies $z$ invariance. This approximation is valid for short propagation lengths, such that the pump does not deplete considerably. The results presented in this paper are based on this approximation. 

For longer propagation lengths the pump does deplete, which reduces the Rabi frequency $\Omega_{\text p}$ and leads to a further reduction in the group velocity. In this case, one can treat the whole waveguide as a series of concatenated short sections along the propagation direction, such that the pump intensity in each section is approximately constant. This treatment allows one to determine the group velocity of the signal in each section and calculate the overall delay of the signal in the whole waveguide.

Inside the waveguide core, the pump field expression is given by
\begin{equation}
{\bm E}_{\text{p}}(r)=A\left(J_{0}(\kappa_{\text{p}}r)\,\hat{\bm z}-i\frac{\tilde{\beta}_{\text{p}}}{\kappa_{\text{p}}^{2}}
J_{1}(\kappa_{\text{p}}r)\,\hat{\bm r}\right),
\end{equation}
with $\hat{\bm z}$ and $\hat{\bm r}$ unit vectors in the propagation and radial directions, respectively; $A$ a constant that determines the amplitude of the pump, $\tilde{\beta}_{\text{p}}$ the complex propagation constant for the pump mode, and
\begin{equation}
\kappa_{\text{p}}:=\sqrt{\omega_{\text{es}}^{2}\epsilon_{\text d}\mu_{\text d}-\tilde{\beta}_{p}^{2}}.
\end{equation}
We define $\eta:=Ad$, which is a parameter adjusted by the amplitude of the pump, with $d=\bm d \cdot\hat{\bm z}=\bm d \cdot\hat{\bm r}$ such that
\begin{equation}
\Omega_{\text p}(r)=\frac{\eta}{\hbar}\left(J_{0}(\kappa_{\text{p}}r)-i\frac{\tilde{\beta}_{\text{p}}}{\kappa_{\text{p}}^{2}}
J_{1}(\kappa_{\text{p}}r)\right).\label{rabifreqfunc}
\end{equation} 

\begin{figure}[t,b] 
       \centering
	\includegraphics[width=0.45\textwidth]{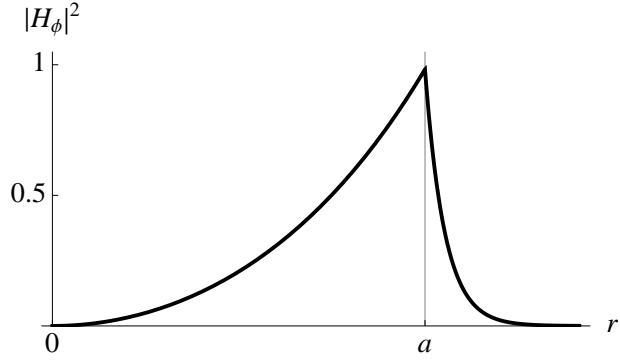}
	\caption{A plot of $|H_{\phi}|^{2}$ of a surface mode showing the transverse profile in arbitrary units. The thin vertical line indicates the interface between the core and cladding regions.}
     \label{fig:surfacemode}   
\end{figure}

The permittivity of the doped core depends on the pump field. The pump field is in a TM surface mode (see Fig.~\ref{fig:surfacemode}),
so the intensity varies steeply in the radial direction; thus the signal field experiences a steeply graded refractive index in the core. As a result, within the core, the scalar wave equation used to solve for the modes of the undoped guide is no longer valid for the modes of the doped guide. In the cladding region, however, the permittivity and permeability are not spatially dependent, so we can still find the fields in this region using the scalar wave equation. 

To solve for the propagation constants and the fields in the doped guide, we must begin with the vector wave equation
\begin{equation}
{\nabla}^{2}{\bm E}+{\bm\nabla}\left(\frac{1}{\epsilon_{\text{eff}}(\omega,r)}{\bm E}\cdot{\bm\nabla}\epsilon_{\text{eff}}(\omega,r)\right)+k_{0}^{2}\epsilon_{\text{eff}}(\omega,r){\bm E}=0,\label{vectorwave}
\end{equation}
with ${\nabla}^{2}$ the Laplacian and
\begin{equation}
\epsilon_{\text{eff}}(\omega,r)=\left\{
\begin{array}{cl}
\epsilon_{0}\left[1+\chi_{\text{d}}+\chi(\omega, r)\right],&r<a,\\
\epsilon_{\text{MM}}(\omega),&r\geq a,
\end{array}
\right.
\end{equation}
the effective permittivity of the waveguide with three-level $\Lambda$ atoms embedded in the core.
As we are restricting our analysis to TM modes for both the pump and signal fields, Eq.~(\ref{vectorwave}) can be decoupled into scalar equations with the $z$ component taking the form
\begin{equation}
\frac{\text{d}^{2}}{\text{d}r^{2}}E_{z}+\left(\frac{1}{r}-\frac{\tilde{\beta}^{2}}{\kappa_{\text{eff}}^{2}(\omega,r)\epsilon_{\text{eff}}(\omega,r)}\frac{\text{d}}{\text{d}r}\epsilon_{\text{eff}}(\omega,r)\right)\frac{\text{d}}{\text{d}r}E_{z}+\kappa_{\text{eff}}^{2}(\omega,r)E_{z}=0\label{TMvectwave},
\end{equation}
with
\begin{equation}
\kappa_{\text{eff}}(\omega,r):=\left\{
\begin{array}{cl}
\sqrt{\omega^{2}\mu_{0}\epsilon_{\text{eff}}(\omega,r)-\tilde{\beta}^{2}},&r<a,\\
i\gamma,&r\geq a.
\end{array}
\right.
\end{equation}
Equation~(\ref{TMvectwave}) is not generally analytically solvable, except at certain frequencies, so we numerically determine the dispersion and attenuation of the low-loss surface mode near the EIT resonance frequency.


\section{Method and Results}\label{results}
There are two main problems with solving Eq.~(\ref{TMvectwave}). First, $\tilde{\beta}$ is complex, meaning it has two free parameters so a standard shooting method cannot be employed~\cite{Press:2007}. The second problem is that the fitness function,
\begin{equation}
f=\left|\frac{H_{\phi,\text{core}}-H_{\phi,\text{clad}}}{H_{\phi,\text{clad}}}\right|,
\end{equation}
with $H_{\phi,\text{core}}$ and $H_{\phi,\text{clad}}$ the $\phi$ component of the magnetic field in the core and cladding regions, respectively, is not convex, so implementing a hill descent algorithm is not sufficient to determine if a particular value of $\tilde{\beta}$ is a solution. To circumvent these problems we draw inspiration from the shooting method and employ a trial method to solve Eq.~(\ref{TMvectwave}) for $E_{z}(r,\omega)$. 

We need to find $E_{z}(r,\omega)$ for a range of frequnecies of interest. We begin at a frequency where an exact solution for a TM surface mode is found. To find the solution at an adjacent frequency, we slightly deviate the known $\tilde{\beta}$ at the former frequency to obtain a trial value for $\tilde{\beta}$ and numerically solve Eq.~(\ref{TMvectwave}) for $E_{z}(r, \omega)$; other components of the field are calculated from this trial $E_{z}(r,\omega)$. To test whether the trial $\tilde{\beta}$ is acceptable, we check if the corresponding trial $E_{z}(r,\omega)$ satisfies the boundary condition at the core-cladding interface to within a tolerance. Mathematically, the acceptance criterion is $f<10^{-9}$. If the trial $E_{z}(r,\omega)$ does not satisfy the boundary condition, a different trial $\tilde{\beta}$ is chosen by slightly deviating the known $\tilde{\beta}$ at the former frequency and the search is repeated. When an acceptable $\tilde{\beta}$ is found, we repeat the search for the next frequency until $E_{z}(r,\omega)$ of the TM surface mode is determined for a range of frequencies of interest.

We apply the same metamaterial parameters presented in our previous work~\cite{Lavoie:2012} to our calculations in this work. Hence, the TM surface mode of the undoped guide found in~\cite{Lavoie:2012} is the mode for the pump of our scheme. The metamaterial parameters are shown in Table~\ref{MMparams}. The core parameters are $\epsilon_{\text d}=1.3\epsilon_{0}$, $\mu_{\text d}=\mu_{0}$ and the core radius $a=4\pi c/\omega_{\text e}$. For the three-level atoms embedded in the core, we set the number density $\rho_{\text a}=1.26\times10^{21}\text{m}^{-3}$ so that the resulting permittivity remains positive. The decay rate of the excited state is $\gamma_{\text{eg}}=10^{5}\text{s}^{-1}$ and that of the hyperfine state is $\gamma_{\text{sg}}=10^{-2}\text{s}^{-1}$. These decay rates are consistent with observed values for $\text{Pr}^{3+}$ ions embedded in bulk Y$_{2}$SiO$_{5}$ crystal and under cryogenic conditions~\cite{Turukhin:2001}. The mean signal frequency is $\omega_\text{s}=\omega_\text{eg}= 0.409\omega_{\text e}$ and the mean pump frequency is $\omega_\text{p}=\omega_\text{es}=0.41\omega_{\text e}$. The group velocity of the pump mode $v_{\text p}=0.47c$.

\begin{table}[h,t,b]
\caption{Parameter values for the metamaterial cladding.}
\begin{tabular}{cc}
\hline Parameter &Value\\
\hline\hline
$\epsilon_{\text b}$&1\\
$\mu_{\text b}$&1\\
$\omega_{\rm e}$&$1.37\times10^{16}{\rm s}^{-1}$\\
$\Gamma_{\rm e}$&$2.73\times10^{13}{\rm s}^{-1}$\\
$\Gamma_{\rm{m}}$&$\Gamma_{\rm e}$\\
$\omega_0$&$0.2\omega_{\rm e}$\\
$F$&$0.5$\\
\hline
\end{tabular}
\label{MMparams}
\end{table}

\begin{figure}[t,b] 
       \centering
	\includegraphics[width=0.6\textwidth]{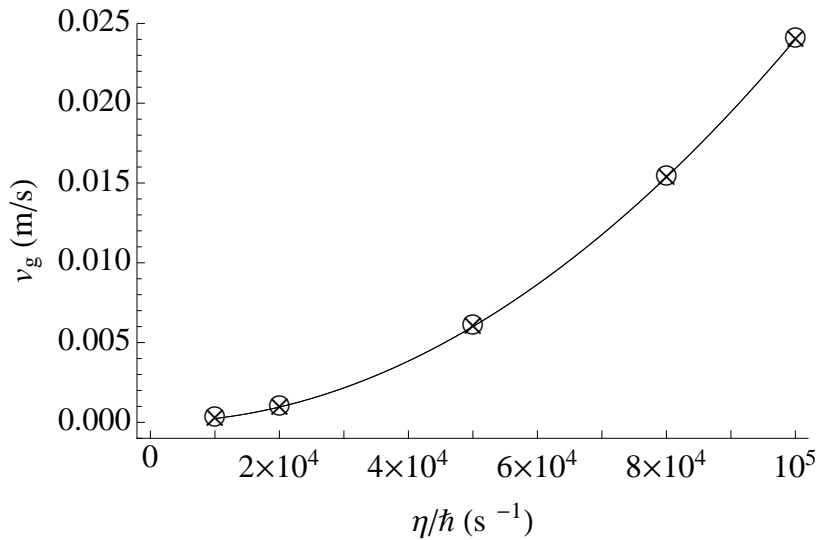}
	\caption{A plot of the numerically calculated group velocity at $\Delta=0$ as a function of pump intensity for a metamaterial-clad guide ($\bigcirc$) with $\Gamma_{\text m}=\Gamma_{\text e}/3$ and a metal-clad guide ($\times$). The solid curve is a least squares quadratic fit.}
     \label{fig:vgplot}   
\end{figure}

Using the method and parameters outlined above, we have calculated the propagation constants of the signal and pump TM surface modes of our scheme. With the complex propagation constant $\tilde{\beta}=\beta+i\alpha$ for a range of frequencies, we have calculated the group velocity of the signal field using $v_{\text{g}}=(\text{d}\beta/\text{d}\omega)^{-1}$. Figure~\ref{fig:vgplot} shows the group velocity of the signal field surface mode in both the metal- and metamaterial-clad waveguides for a range of $\eta$. The plot shows that the group velocity of the signal for a metamaterial-clad guide is proportional to $\eta^{2}$, and as $\eta\propto|\Omega(r)|$, this agrees with EIT theory~\cite{Lambropoulos:2007}.

\begin{figure}[t,b] 
       \centering
	\includegraphics[width=0.6\textwidth]{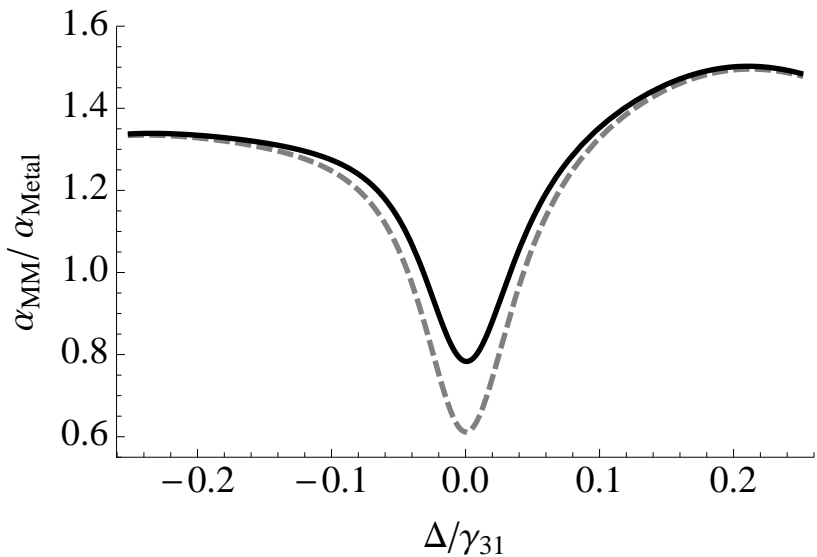}
	\caption{Plots of the relative attenuation of the surface mode of the metamaterial guide to that of a metal guide ($\alpha_{\text{MM}}/\alpha_{\text{Metal}}$) with $\eta/\hbar=5\times10^{4}\text{s}^{-1}$, for $\Gamma_{\text{m}}=\Gamma_{\text{e}}/3$ (solid) and $\Gamma_{\text{m}}=\Gamma_{\text{e}}/100$ (dashed).}
     \label{fig:lowlosseit}   
\end{figure}

Using the imaginary part of the propagation constant, we have obtained the attenuation curves for both the metamaterial- and metal-clad waveguides as a function of detuning $\Delta$ as shown in Fig.~\ref{fig:lowlosseit}. At the EIT resonance (i.\e.\ zero detuning), the metamaterial-dielectric guide provides nearly a 20\% reduction in attenuation over a metal-dielectric guide for $\Gamma_{\text{m}}=\Gamma_{\text{e}}/3$ (solid line). For $\Gamma_{\text{m}}=\Gamma_{\text{e}}/100$ (dashed line), the attenuation reduction improves to near 40\% over a metal-dielectric guide. When $\Gamma_{\text{m}}=\Gamma_{\text{e}}$ the attenuation of a metamaterial-dielectric guide is comparable to a metal-dielectric guide. For layered structures, such as the fishnet design, previous work~\cite{Penciu:2010} has shown that $\Gamma_{\text{m}}\leq\Gamma_{\text{e}}$ and a value of $\Gamma_{\text{m}}=\Gamma_{\text{e}}/3$ should be feasible with current metamaterial technology~\cite{Zhou:2008}. Hence, our cylindrical metamaterial-dielectric waveguide is capable of supporting a low-loss TM surface mode.

\begin{figure}[t,b] 
       \centering
	\includegraphics[width=0.6\textwidth]{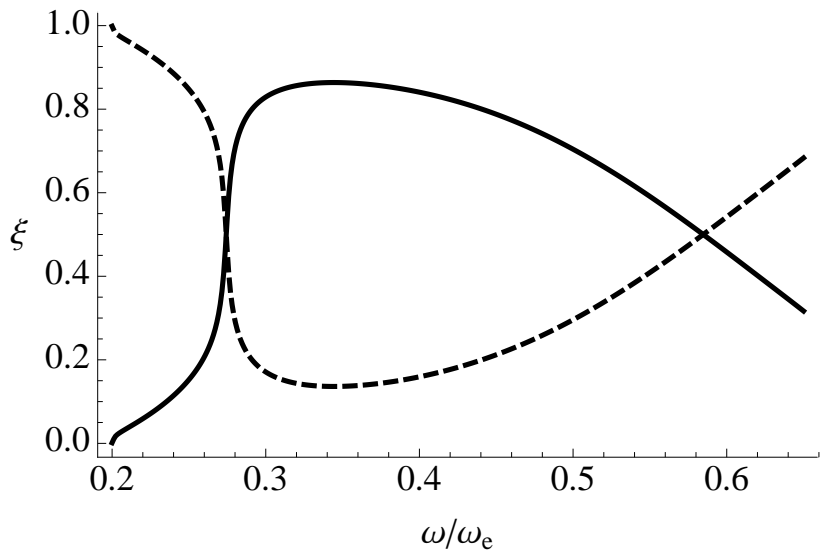}
	\caption{Plots of the fraction of the total energy in the core (solid) and the cladding (dashed) as a function of frequency for the low-loss surface mode of the cylindrical metamaterial-dielectric waveguide.}
     \label{fig:fracen}   
\end{figure}

To understand why the attenuation of the low-loss surface mode does not approach zero near $\omega_{\text s}$, as it does for the flat interface~\cite{Kamli:2008}, we plot the fraction of the total energy of the mode that resides in both the core and cladding. Figure~\ref{fig:fracen} shows the fractional energy $\xi$ of the TM surface mode of the cylindrical metamaterial-dielectric guide as a function of frequency. For a given mode the fractional energy is the ratio of the energy in one part of the guide, the core for instance, to the total energy of the mode. The solid line is for the fractional energy in the core whereas the dashed line is for the cladding. The fraction of energy in the dielectric does not reach 1, and some of the signal field penetrates into the metamaterial, leading to a nonzero attenuation.

The results above show that embedding three-level $\Lambda$ atoms in the core of a metamaterial-dielectric waveguide provides coherently controllable slowing of the signal field with reduced losses. In the next section, we discuss the loss mechanism in the metamaterial waveguide and explain how the existence of the low loss TM surface mode is due to $\Gamma_{\text{m}}$ taking a value less than $\Gamma_{\text{e}}$ . 
 

\section{Discussion}\label{discussion}

In our calculation, we have treated the claddings as if they are infinitely thick. In practice, the TM surface modes for the pump and signal fields only penetrate into the metamaterial to some finite depth, so a finite thickness of the cladding regions is sufficient. The skin depth for both the signal and pump fields is about $0.1\lambda_{\text s}$, with $\lambda_{\text s}$ the wavelength of the signal at the EIT resonance. As typical fishnet metamaterial has thickness of $0.1\lambda_{\text s}$~\cite{Boltasseva:2008}, a few layers of fishnet metamaterial (with dielectric spacers in between) is sufficient for the cladding.

We find that the attenuation dip near $\omega_{\text s}$ (Fig.~\ref{fig:lowlosseit}) is not as pronounced in the presence of transverse field confinement in the cylindrical waveguide. The flat metamaterial-dielectric interface has near-zero attenuation near this frequency, whereas the attenuation dip in the cylindrical guide does not approach zero. The cause for the nonzero attenuation is that the transverse confinement limits the amount of electromagnetic energy that is ejected from the lossy metamaterial.

To illustrate why transverse confinement lessens the attenuation dip near $\omega_{\text s}$, consider a cylindrical guide in the limit of an infinite radius. In this limit, the interface that the surface wave propagates along becomes flat and the dispersion relation for the cylindrical waveguide reduces to that for a flat interface waveguide. In the case of the flat interface, the energy can reside almost completely on the dielectric side, leading to near-zero loss~\cite{Moiseev:2010}. Alternatively, for a cylindrical guide in the zero core radius limit, the core almost vanishes; hence, the energy propagates almost entirely in the lossy metamaterial, which leads to significant loss. For a waveguide with a finite radius, the fraction of the total energy that can be confined to the dielectric lies between 0 and 1; thus there is a non-negligible amount of attenuation due to the field interacting with the metamaterial cladding. 

We have considered fishnet metamaterial for our scheme, but the scheme is not restricted to using fishnet metamaterial. Other types of metamaterial are applicable, as long as they have permittivity and permeability functions resembling Eqs.~(\ref{epsilonMM}) and~(\ref{muMM}), with $\epsilon'_{\text{MM}}<0$, $0<\mu'_{\text{MM}}<1$, and $\Gamma_{\text m}<\Gamma_{\text e}$. Using other metamaterials is of interest, because when layered fishnet metamaterials are bent to form a cylindrical cladding, the metamaterial may not interact with the fields isotropically. There now exist various promising fabrication techniques to building bulk metamaterials instead of layered ones~\cite{Boltasseva:2008,Vignolini:2012}, which are isotropic.

This system is assumed to operate at a cryogenic temperature as it provides the advantage of mitigating decoherence effects that reduce EIT efficacy~\cite{Ham:1997}. An auxiliary advantage of the cryogenic temperatures is increased charge mobility in metals, thereby decreasing attenuation due to a portion of the electromagnetic fields propagating through the metamaterial cladding~\cite{Singh:2010}.

The major drawback of cryogenic temperatures is this system needs to be contained within a cryostat, which will complicate its integration and use with other systems that need to operate at higher temperatures. We are, however, only suggesting the use of cryogenic temperatures in the interest of achieving a proof-of-principle design. Other effects of cryogenic temperatures to consider are a modified refractive index in the dielectric core and altered core dimensions. These changes, however, can be measured and compensated for based on the operational requirements.

Though our main result concerns controllable slow light, the use of EIT in this system means it is also capable of stopping the signal pulse completely by turning off the control field entirely~\cite{Turukhin:2001}. An important application of stopped light is optical quantum memories~\cite{Lvovsky:2009}, to which EIT is well suited due to its capacity for direct control. However, the narrow bandwidth restriction inherent in EIT-based schemes places a lower-bound restriction on the duration of the signal pulse. This has implications in terms of data rates for quantum communication and quantum computation schemes that would use this memory.

The imposed lower bound on pulse duration, which correlates directly with a minimum spatial extent, sets an upper limit on the pulse repetition rate. As the maximum allowable overlap for pulses to remain distinguishable restricts the minimum temporal separation between pulses, the number of pulses that can be processed within a given time window is limited. The effects of the narrow bandwidth on the information storage capability could be avoided, however, by storing information in ways that are not affected by the reduced bandwidth, such as amplitude and phase~\cite{Liu:2001,Appel:2008}.

The operational bandwidth of this device can easily be broadened by increasing the pump intensity, but comes at the expense of the group velocity reduction that can be achieved. This is because the spectral width of the EIT transparency window is related to the intensity of the pump field~\cite{Lambropoulos:2007}. Alternatively, the bandwidth can be broadened by using another technique to achieve slow light, such as Raman amplification~\cite{Sharping:2005} or photon echo techniques~\cite{Saglamyurek:2010}. However, such schemes come at the expense of increased group velocities and controllability. The Raman amplification technique does have the benefit of room temperature operation.



\section{Summary}\label{summary}

We have considered two cylindrical dielectric-core waveguides of equal size and dimensions, with one having a metal cladding and the other a fishnet metamaterial cladding; other types of metamaterial cladding are also considerable. The permittivities of both claddings are given by the Drude model, but the metal cladding has constant permeability and the metamaterial cladding is given by the modified-Drude model. The signal field propagating through the waveguides is controllably slowed via electromagnetically induced transparency (EIT)  due to the three-level $\Lambda$ atoms homogeneously doped throughout the dielectric core of each waveguide. We have shown that the low-loss TM surface mode exists in the metamaterial waveguide provided that the permeability of the metamaterial cladding has a real part less than unity, i.\e.\ $\Re{\mu}<1$, and the magnetic damping rate is less than the electric damping rate, i.\e.\ $\Gamma_{\text{m}}<\Gamma_{\text{e}}$. Previous work has shown that this condition is always true for certain optical metamaterials~\cite{Penciu:2010}. We predict that  such a cylindrical metamaterial-clad waveguide is capable of delivering the same slowing of light but with reduced loss compared to the cylindrical metal-clad waveguide. 

\begin{acknowledgments}
We appreciate financial support form AITF and NSERC, and BCS is partially supported by a CIFAR Senior Fellowship.
\end{acknowledgments}

\end{document}